\documentstyle[aps,twocolumn]{revtex}
\newcommand{\grad}{\nabla} 
\newcommand{\be}{\begin{eqnarray}}
\newcommand{\ee}{\end{eqnarray}}
\newcommand{\OP}{\vec{\psi}}
\newcommand{\SG}{\vec{\sigma}}
\newcommand{\M}{\vec{m}}
\newcommand{\U}{\vec{u}}
\newcommand{\F}{{\cal F}}
\newcommand{\A}{\omega_{0}}
\newcommand{\B}{\omega_{2}}
\newcommand{\C}{\omega_{1}}
\newcommand{\T}{\vec{\Theta}}
\renewcommand{\baselinestretch}{1.06}
\begin{document}

\title{Spinodal Decomposition and the Tomita Sum Rule}

\author{Gene F. Mazenko }
\address{The James Franck Institute and the Department of Physics \\
         The University of Chicago \\
         Chicago, Illinois 60637 }
\date{\today}
\maketitle
%
%
\begin{abstract}

The scaling properties of a
phase-ordering system with a conserved order parameter are
studied.  The theory developed leads to scaling functions 
satisfying certain general properties including
the Tomita sum rule.  The theory also 
gives good agreement with
numerical results for the order parameter scaling function
in three dimensions.
The values of the associated nonequilibrium decay exponents
are given by the known lower bounds.

\end{abstract}
\draft
\pacs{PACS numbers: 05.70.Ln, 64.60.Cn, 64.75.+g, 98.80.Cq}

\section{INTRODUCTION}

In the area of phase-ordering kinetics\cite{221} there are a wide variety
of systems which satisfy a form of dynamical scaling.  
In the case of systems with a nonconserved order parameter (NCOP) we have a 
simple approximate theoretical model due to 
Ohta, Jasnow and Kawasaki\cite{OJK} (OJK)
which captures the main scaling properties of the 
associated physical systems.  In the case of a conserved order
parameter (COP) the situation is much less satisfactory.  
The difficulty in the COP case is that
there are competing length scales which lead to the necessity of 
treating crossover.  
This crossover connects
up the short-scaled-distance nonanalytic domain wall behavior associated
with Porod's law\cite{Porod0,DAB,Porod} and the large distance constraints of the 
conservation law.  
A theory is offered in this paper which is consistent with  all of the 
prominent scaling features in
the case of a conserved  scalar order parameter.

An auxiliary field method is used 
in essentially all the available 
explicit calculations of scaling 
functions in phase-ordering kinetics.
Thus in the OJK approach\cite{OJK,OP,YJ,PR,BH},
and our previous work\cite{TUG,TUGCOP,GFMCOP},
a local mapping from the original order parameter $\psi$ onto an
approximately gaussian variable $m$ was developed.
While these methods can be shown to work well for the NCOP case,
there are severe limitations in the COP case.  In particular 
Yeung, Oono and Shinozaki\cite{YOS} found that such a local mapping in the
COP case led to mathematically unacceptable results within the
theory.  All of these theories were developed with the idea that
the mapping, $\psi \rightarrow\psi (m)$, leads to an equation of motion for
$m$ which could be argued to be approximately consistent with a gaussian distribution
for $m$.  In Refs.(\onlinecite{EXP}) and (\onlinecite{EXP2}) a different approach was taken
in treating the NCOP case.  It was shown that the equation of motion for
the auxiliary field $m$ can be constructed with appeal only to the form
of the growth law, defined below, and certain general symmetry constraints.  Using a
novel expansion method it was shown how one could obtain the
OJK result in zeroth order of a systematic expansion.  Higher order corrections for 
the associated nonequilibrium exponents were also obtained. 

In this paper this general idea is applied to the simplest COP
system given by the Cahn-Hilliard\cite{CH} (CH) model.
One is led to introduce a nonlocal mapping between 
the order parameter and the auxiliary field and only certain general
constraints on the theory, like the conservation law and the 
generalized form of the Tomita
sum rule\cite{Tom}, are used to determine the parameters characterizing the
nonlocal mapping and the correlation function for the underlying auxiliary
field.  Thus one has a nonlinear selection problem where one
simultaneously constructs the parameters of the mapping and the 
scaling function.  The selected scaling function
is found to be in good agreement with the best
available numerical results in three dimensions.  The nonequilibrium exponents are
also determined in this theory and take on values, as for the OJK
theory, corresponding to the known\cite{YRD} lower bounds.

\section{Phenomenology}

The equation of motion in the Cahn-Hilliard model governing the 
conserved scalar order parameter, 
$\psi$, is given in dimensionless form 
by:
\be
\dot{\psi}=\nabla^{2}\left(V'(\psi)-\nabla^{2}\psi\right)
~~~.
\label{eq:9}
\ee
$V(\psi)$ is a degenerate double-well potential.  
Typically this system is driven by a set of
uncorrelated random initial conditions.
We look here at critical quenches
where $<\psi >=0$.
In this case the
system is unstable and responds by locally growing competing degenerate
patches of the stable 
low-temperature-ordered phases.
These patches correspond to domains
separated by sharp walls of width $\xi$. As time evolves these domains coarsen
and the order grows to progressively longer length scales.
The growth law, $L(t)$, increases without bound with time $t$ after the
quench. At long enough times
$L(t)$ dominates, $L(t)\gg\xi$, and
the order parameter
correlation function satisfies the scaling equation
\cite{MLK,BSsc,Furu78}
\be
C({\bf r},t)
\equiv\langle\psi({\bf r},t)\psi (0,t)\rangle
=\psi_{0}^{2}F(x)
\label{eq:3}
\ee
where $x\equiv r/L(t)$,  and $\psi _0$ is the magnitude
of $\psi$ in the ordered state. The structure factor, the Fourier
transform of $C({\bf r},t)$, satisfies
\be
C({\bf q},t)=L^{d}\psi_0^2 F(Q)
~~~,
\ee
where $Q\equiv qL$ is a scaled wavenumber and $d$ the number of spatial dimensions.
There are a number of general properties a correct theory
of the phase ordering CH model 
must satisfy\cite{SO,GFMCOP}:

1. The growth law is given by the Lifshitz-Slyozov-Wagner\cite{LSW} form:
$L\approx t^{1/3}$. This result can be obtained 
a number of different ways\cite{LT,BRGL,BLSA} that are all consistent.

2.  The scaled-structure factor $F(Q)$ has a $Q^{4}$  behavior\cite{Yeung} for 
small scaled wavenumbers $Q$
directly reflecting the conservation law. 

3.  The scaled-structure factor satisfies  Porod's law\cite{Porod0,DAB,Porod,OP,cs10}
for large
scaled wavenumbers, $F(Q)\approx Q^{-(d+1)}$. 

4.  The scaled-structure factor  also satisfies the Tomita\cite{Tom,OP}
sum rules.  The
scaling function $F(x)$
has no even terms, except the first, in its expansion
in powers of $x$:
\be
F(x)=1+ F_{1}x+F_{3}x^{3}+\ldots
~~~.
\ee

While there are theories which satisfy some of these requirements,
there
has been none so far which satisfies all four.
Our goal here is to find the simplest 
theory for
the COP case  which does satisfy all of these properties and leads
to an explicit form for the scaled structure factor which can be
compared to the best numerical determinations of $F(Q)$.
The challenge is to 
match the small
$Q^{4}$ behavior for $F(Q)$ with the large $Q^{-4}$ behavior (in three
dimensions) while preserving the
Tomita sum rule.  Our approach will be similar to that developed in
Refs.(\onlinecite{EXP}) and (\onlinecite{EXP2}) but with some significant 
differences.  

\section{Auxiliary Field mapping for COP}

As in previous work, we assume in the scaling regime that the order 
parameter can be decomposed into 
an ordering component, which contributes to the order parameter structure
factor at ${\cal O}(1)$, and a fluctuating piece which is of higher order in 
powers of $1/L(t)$:
\be
\psi =\tilde{\psi}+\Theta/L 
\label{eq:10}
\ee
where $\Theta$ is ${\cal O}(1)$.
Here we assume that the  ordering component, $\tilde{\psi}$,
can be written in the form:
\be \tilde{\psi}=\sigma (m+u)
\label{eq:12}
~~~,
\ee
where, as usual, $\sigma (m)$ is the solution of the classical 
interfacial equation
\be
\frac{d^{2}\sigma}{dm^{2}}=V'(\sigma) ~~~,
\ee
with the boundary conditions 
$\lim_{m\rightarrow\pm \infty}\sigma (m)=\pm \psi_{0}$.  It turns
out in this case that it is necessary to introduce two independent
fields  $m$ and $u$.  We will organize the
theory such that $m$ is treated as the auxiliary field which,
as in the NCOP case, governs the short scaled distance physics
associated with Porod's law and the Tomita sum rule.  We will
assume, as a first approximation, that $m$ is a gaussian variable
driven by an equation of motion of a general form compatible with
scaling and with coefficients determined by selection processes
described in detail below.
The quantity $\Theta$ is, in principle, to be
determined as a function of the more fundamental variables
$m$ and $u$. In practice we will not need to construct the
correlations for the field $u$ explicitly or assume that it is
a gaussian field.

If we insert the ansatz given by Eq.(\ref{eq:10}) into the equation of motion,
 Eq.(\ref{eq:9}),  we obtain
\be
\frac{\partial \tilde{\psi}}{\partial t} +\frac{\partial}{\partial t}
\left(\frac{\Theta}{L}\right)
=\nabla^{2}\left(V'(\tilde{\psi}+\Theta/L)
-\nabla^{2}\left(\tilde{\psi}+\Theta/L\right)\right)
~.
\label{eq:8}
\ee
We can use scaling arguments to estimate the contributions of
various terms.  We explicitly assume that $L(t)=L_{0}t^{1/3}$
in the scaling regime.  The first term on the left-hand side of
Eq.(\ref{eq:8}),
$\frac{\partial \tilde{\psi}}{\partial t}$,
is of ${\cal O} (L^{-3})$ in the scaling regime.  Then we can estimate
\be
\frac{\partial}{\partial t}\left(\frac{\Theta}{L}\right)\approx {\cal O}(L^{-4})
\ee
and this term can be dropped when compared to the leading order
in the equation of motion.  Next we can expand
\be
V'(\tilde{\psi}+\Theta/L)=V'(\tilde{\psi})+V''(\tilde{\psi})\frac{\Theta}{L}
+{\cal O}(L^{-2})
~~~.
\ee
In comparing these two terms we have, using Eq.(\ref{eq:12}) and
$m\approx L$, that
\be
V'(\tilde{\psi})=\frac{d^{2}\sigma (m+u)}{dm^{2}}\approx {\cal O}(L^{-2})
\ee
and the term proportional to
$V''(\tilde{\psi})=V''(\psi_{0})+{\cal O}(L^{-1})$
dominates at leading order for long times.  Finally we have that the last
term on the right hand side of Eq.(\ref{eq:9}),
\be
\nabla^{4}\left(\tilde{\psi}+\Theta/L\right)\approx {\cal O}(L^{-4})
~~~,
\ee
and can be dropped in Eq.(\ref{eq:8}).  The equation of motion
then reduces to the key
result
\be
\frac{\partial \tilde{\psi}}{\partial t}=\frac{\kappa_{0}}{L}\nabla^{2}\Theta ~~~.
\label{eq:20}
\ee
where $\kappa_{0}\equiv V''(\psi_{0}) >0$.

The quantity $\Theta$ is arbitrary except for the very important
constraints that it be of ${\cal O}(1)$ in the scaling regime and it must
be consistent with the system ordering.
It is at this stage that we realize that there is an apparent flexibility
in the construction of the scaling solution.
Since there is a belief that the scaling functions are
universal, there must be mechanisms, like the nonlinear eigenvalue problem
encountered in Ref.(\onlinecite{TUG},  which selects the scaling structures
which do not directly depend on the physics at the smaller length scales. 
The key assumption in going forward is that the auxiliary field method 
can be used
to describe the {\it inner} scaling regime and there is a crossover to
an {\it outer} scaling regime dominated by the conservation law.
The building blocks we can use to construct $\Theta$, which are compatible with
this crossover and  are of ${\cal O}(1)$, are
$\tilde{\psi}$ and  $\sigma (m)$.

The simplest assumption\cite{few} for the Fourier transform $\Theta_{q}(t)$, 
robust enough
to give a satisfactory solution to the problem, is of the form:
\be
\Theta_{q}(t)=\tilde{M}_{q}(t)\tilde{\psi}_{q}(t)-\tilde{N}_{q}(t)\sigma_{q}(t)
\ee
where $\tilde{M}_{q}(t)$ and $\tilde{N}_{q}(t)$ are functions to be
determined
The equation of motion, Eq.(\ref{eq:20}), is given then by:
\be
\frac{\partial\tilde{\psi}_{q}(t)}{\partial t}=
-M_{q}(t)\tilde{\psi}_{q}(t)+N_{q}(t)\sigma_{q}(t)
\label{eq:27}
\ee
where
\be
M_{q}(t)=\frac{\kappa_{0}q^{2}}{L(t)}\tilde{M}_{q}(t)
\ee
\be
N_{q}(t)=\frac{\kappa_{0}q^{2}}{L(t)}\tilde{N}_{q}(t)
~~~.
\ee 
We can easily obtain a partial solution to Eq.(\ref{eq:27}).  Let
us define the auxiliary quantity
\be
U_{q}(t_{1},t_{2})=
e^{\left(-\int_{t_{2}}^{t_{1}}d\tau~M_{q}(\tau )\right)}
\label{eq:19}
\ee
and write
\be
\tilde{\psi}_{q}(t)=U_{q}(t,t_{0})\chi_{q}(t)
~~~.
\ee
Taking the time derivative of this expression we find
\be
\frac{\partial\tilde{\psi}_{q}(t)}{\partial t}=
-M_{q}(t)\tilde{\psi}_{q}(t)
+U_{q}(t,t_{0}) \frac{\partial\chi_{q}(t)}{\partial t}
~~~.
\label{eq:28}
\ee
Comparing Eqs.(\ref{eq:27}) and (\ref{eq:28}), we obtain the 
equation for $\chi_{q}(t)$,
\be
U_{q}(t,t_{0}) \frac{\partial\chi_{q}(t)}{\partial t}
=N_{q}(t)\sigma_{q}(t) ~~~,
\ee
which can be rewritten as
\be
 \frac{\partial\chi_{q}(t)}{\partial t}=
U_{q}(t_{0},t)N_{q}(t)\sigma_{q}(t)
~~~.
\ee
This equation has the solution
\be
\chi_{q}(t)=\tilde{\psi}_{q}(t_{0})
+\int^{t}_{t_{0}}d\bar{t}~U_{q}(t_{0},\bar{t})N_{q}(\bar{t})\sigma_{q}(\bar{t})
\ee
or
\be
\tilde{\psi}_{q}(t)=U_{q}(t,t_{0})\tilde{\psi}_{q}(t_{0})
+\int^{t}_{t_{0}}d\bar{t}~U_{q}(t,\bar{t})
N_{q}(\bar{t})\sigma_{q}(\bar{t})
~~~.
\label{eq:29}
\ee
This nonlocal relationship between $\tilde{\psi}_{q}(t)$ and $\sigma_{q}(t)$ should be
contrasted with the local mapping used in previous theories.
Because of the nonlocality the
criticism of local mappings in the COP case  due to 
Yeung, Oono, and Shinozaki\cite{YOS,NL}  is irrelevant for the discussion
here.

Notice that we need to determined the functions 
$M_{q}(t)$, $N_{q}(t)$, and the variance of the field $m$.  
Averages over $m$ are discussed in more detail below.  
Focusing on $N$ and $M$,
for our purposes here,
we only need these
quantities in the scaling regime.  Inspection of Eq.(\ref{eq:19}) shows
that a general form compatible with
scaling is given by
\be
M_{q}(t)=\frac{\partial G(Q^{2})}{\partial t}
\ee
where $Q=qL(t)$.  Similarly,
\be
N_{q}(t)=\frac{\partial G_{0}(Q^{2})}{\partial t}
~~~.
\ee
We can then write
\be
M_{q}(t)=\frac{\partial G(Q^{2})}{\partial Q^{2}}\frac{\partial Q^{2}}{\partial t}
=\frac{\partial G(Q^{2})}{\partial Q^{2}}
\frac{\partial }{\partial t}q^{2}L_{0}^{2}t^{2/3}
\nonumber
\ee
\be
=\frac{\partial G(Q^{2})}{\partial Q^{2}}\frac{2 }{3}\frac{Q^{2} }{ t}
\equiv \frac{2 }{3}\frac{H(Q^{2}) }{ t}
\label{eq:28a}
\ee
where
\be
H(Q^{2})=Q^{2}\frac{\partial G(Q^{2})}{\partial Q^{2}}
~~~.
\ee
Similarly,
\be
N_{q}(t)=\frac{2 }{3}\frac{H_{0}(Q^{2}) }{ t}
~~~.
\ee
For our purposes it will be sufficient to assume that
$H$ and $H_{0}$ have power series forms:
\be
H(Q)=\sum_{n=2}\gamma_{n}Q^{n}
\label{eq:31}
\ee
\be
H_{0}(Q)=\sum_{n=2}\gamma_{n}^{0}Q^{n}
~~~.
\ee
For reasons discussed below, we will work explicitly with a model where only
$\gamma_{2}$, $\gamma_{10}$, $\gamma_{2}^{0}$, and $\gamma_{10}^{0}$ are nonzero. 

\section{Structure Factor}

The quantity of central interest is the order parameter
structure factor:
\be
C(q,t_{1},t_{2})=<\psi_{q}(t_{1})\psi_{-q}(t_{2})>
\nonumber
\ee
\be
=<\tilde{\psi}_{q}(t_{1})\tilde{\psi}_{-q}(t_{2})>
~~~,
\ee
where in the second line we recognize that in the scaling regime only
the ordering component of the order parameter contributes to the
structure factor.  Inserting Eq.(\ref{eq:29}) for $\tilde{\psi}_{q}(t_{1})$
we obtain:
\be
C(q,t_{1},t_{2})=
<\int_{t_{0}}^{t_{1}} ~d\bar{t}_{1}~U_{q}(t_{1},\bar{t}_{1})
N_{q}(\bar{t}_{1})\sigma_{q}(\bar{t}_{1})
\nonumber
\ee
\be
\times\int_{t_{0}}^{t_{2}} ~d\bar{t}_{2}~U_{q}(t_{2},\bar{t}_{2})
N_{q}(\bar{t}_{2})\sigma_{q}(\bar{t}_{2})>
\nonumber
\ee
\be
=\int_{t_{0}}^{t_{1}} ~d\bar{t}_{1}~U_{q}(t_{1},\bar{t}_{1})
N_{q}(\bar{t}_{1})
\nonumber
\ee
\be
\times\int_{t_{0}}^{t_{2}} ~d\bar{t}_{2}
~U_{q}(t_{2},\bar{t}_{2})
N_{q}(\bar{t}_{2})
C_{\sigma}(q,\bar{t}_{1},\bar{t}_{2})
\label{eq:44}
\ee
where
\be
C_{\sigma}(q,t_{1},t_{2})=<\sigma_{q}(t_{1})\sigma_{q}(t_{2})>
~~~.
\ee
Notice that we have assumed that the correlations with the initial state
have decayed to zero for long times compared to the ${\cal O}(1)$
terms contributing to the scaling function. 
We also note that $C(q,t_{1},t_{2})$ depends only on the magnitude
of ${\bf q}$.

The next step is to realize that $C_{\sigma}(q,t_{1},t_{2})$, which
we calculate explicitly below,
satisfies a scaling relation:
\be
C_{\sigma}(q,t_{1},t_{2})=L^{d}(t_{1},t_{2})
\psi_{0}^{2}F_{\sigma}(qL(t_{1},t_{2}),t_{1},t_{2})
~~~.
\label{eq:46}
\ee
Inserting Eq.(\ref{eq:46}) into Eq.(\ref{eq:44}) gives
\be
C(q,t_{1},t_{2})=
\int_{t_{0}}^{t_{1}} ~d\bar{t}_{1}~U_{q}(t_{1},\bar{t}_{1})
N_{q}(\bar{t}_{1})
\nonumber
\ee
\be
\times\int_{t_{0}}^{t_{2}} ~d\bar{t}_{2}
~U_{q}(t_{2},\bar{t}_{2})
N_{q}(\bar{t}_{2})L^{d}(\bar{t}_{1},\bar{t}_{2})\psi_{0}^{2}
F_{\sigma}(qL(\bar{t}_{1},\bar{t}_{2}),\bar{t}_{1},\bar{t}_{2})
\label{eq:48}
~.
\ee
Let us now define
\be
L^{2}_{T}(t_{1},t_{2})=\frac{1}{2}\left(L^{2}(t_{1})+L^{2}(t_{2})\right)
~~~,
\label{eq:39}
\ee
and choose a time $T$ such that 
\be
L_{T}(t_{1},t_{2})=L(T)
~~~.
\ee
Remembering  $L(t)=L_{0}t^{1/3}$, we have
\be
T=\left(\frac{1}{2}\left(t_{1}^{2/3}+t_{2}^{2/3}\right)\right)^{3/2}
~~~.
\ee
Clearly for $t_{1}=t_{2}=t$, $T=t$.
Now make the change of variables $\bar{t}_{1}=Ts_{1}$ and $\bar{t}_{2}=Ts_{2}$
in Eq.(\ref{eq:48}).  This requires treatment of the quantity
\be
L_{T}(\bar{t}_{1},\bar{t}_{2})=
L_{T}(Ts_{1},Ts_{2})
\nonumber
\ee
\be
=L(T)\ell(s_{1},s_{2})=L_{T}(t_{1},t_{2})\ell(s_{1},s_{2})
~~~,
\ee
where
\be
\ell(s_{1},s_{2})=\sqrt{\frac{1}{2}(s_{1}^{3/2}+s_{2}^{3/2})}
~~~.
\ee
Eq.(\ref{eq:48}) then takes the form:
\be
C(q,t_{1},t_{2})=\psi_{0}^{2}
\int_{t_{0}/T}^{t_{1}/T} ~Tds_{1}~U_{q}(t_{1},Ts_{1})
N_{q}(Ts_{1})
\nonumber
\ee
\be
\times\int_{t_{0}/T}^{t_{2}/T} ~Tds_{2}
~U_{q}(t_{2},\bar{t}_{2})
N_{q}(Ts_{2})
\nonumber
\ee
\be
\times L^{d}(Ts_{1},Ts_{2})
F_{\sigma}(qL_{T}(t_{1},t_{2})\ell(s_{1},s_{2}),Ts_{1},Ts_{2})
\nonumber
\ee
\be
=\psi_{0}^{2}\int_{t_{0}/T}^{t_{1}/T} ~ds_{1}~U_{q}(t_{1},Ts_{1})
\int_{t_{0}/T}^{t_{2}/T} ~ds_{2}~U_{q}(t_{2},Ts_{2})
\nonumber
\ee
\be
\times TN_{q}(Ts_{1})TN_{q}(Ts_{2})
L^{d}(T)\ell^{d}(s_{1},s_{2})
\nonumber
\ee
\be
\times F_{\sigma}(Q\ell(s_{1},s_{2}),Ts_{1},Ts_{2})
\nonumber
\ee
\be
=L^{d}(T)\psi_{0}^{2}F(Q,t_{1},t_{2})
\ee
where $Q=qL_{T}(t_{1},t_{2})=qL(T)$ and  the physical scaling function is given by
\be
F(Q,t_{1},t_{2})=
\int_{t_{0}/T}^{t_{1}/T} ~ds_{1}~U_{q}(t_{1},Ts_{1})
\int_{t_{0}/T}^{t_{2}/T} ~ds_{2}
\nonumber
\ee
\be
\times U_{q}(t_{2},Ts_{2})
TN_{q}(Ts_{1})TN_{q}(Ts_{2})
\nonumber
\ee
\be
\times \ell^{d}(s_{1},s_{2})
F_{\sigma}(Q\ell(s_{1},s_{2}),Ts_{1},Ts_{2})
~~~.
\ee
We can simplify things a bit in the quantities $U_{q}$,
defined by Eq.(\ref{eq:19}), where
we have in the argument of the exponentials
\be
\int_{Ts_{1}}^{t}~d\tau ~M_{q}(\tau)
=\int_{s_{1}}^{t/T}~dz_{1} ~M_{q}(Tz_{1})
~~~.
\ee
Inserting our general form for $M_{q}$, given by Eq.(\ref{eq:28a}),
we have
\be
\int_{s_{1}}^{t/T}~d\tau ~M_{q}(\tau)
=\int_{s_{1}}^{t/T}~dz_{1} ~\frac{2}{3}\frac{H(qL(Tz_{1}))}{Tz_{1}}
\nonumber
\ee
\be
=\frac{2}{3}\int_{s_{1}}^{t/T}\frac{dz_{1}}{z_{1}}H(Q\ell(z_{1}))
\ee
where
$\ell (z_{1})=z_{1}^{1/3}$,
and similarly
\be
TN_{q}(Ts_{1})=T\frac{2}{3}\frac{H_{0}(Q\ell(s_{1}))}{Ts_{1}}
=\frac{2}{3}\frac{H_{0}(Q\ell(s_{1}))}{s_{1}}
~~~.
\ee
Then 
\be
U_{q}(t_{1},Ts_{1})=e^{\left(-\frac{2}{3}\int_{s_{1}}^{t/T}
\frac{dz_{1}}{z_{1}}H(Q\ell(z_{1}))\right)}
~~~,
\ee
and the scaling function is given by
\be
F(Q,t_{1},t_{2})=
\int_{t_{0}/T}^{t_{1}/T} ~ds_{1}~U_{q}(t_{1},Ts_{1})
\int_{t_{0}/T}^{t_{2}/T} ~ds_{2}~U_{q}(t_{2},Ts_{2})
\nonumber
\ee
\be
\times \frac{2}{3}\frac{H_{0}(Q\ell(s_{1}))}{s_{1}}
\frac{2}{3}\frac{H_{0}(Q\ell(s_{2}))}{s_{2}}
\nonumber
\ee
\be
\times\ell^{d}(s_{1},s_{2})
F_{\sigma}(Q\ell(s_{1},s_{2}),Ts_{1},Ts_{2})
~~~.
\ee
We can then make one final change of integration variables to
$y_{i}=s_{i}^{2/3}$ ($i=1,2$)..
We have then
\be
\ell(s_{1},s_{2})=\left(\frac{s_{1}^{2/3}+s_{2}^{2/3}}{2}\right)^{1/2}
\nonumber
\ee
\be
=\left(\frac{y_{1}+y_{2}}{2}\right)^{1/2}=\tilde{\ell}(y_{1},y_{2})
~~~,
\ee
the equation for the scaling function becomes,
\be
F(Q,t_{1},t_{2})=
\int_{(t_{0}/T)^{2/3}}^{(t_{1}/T)^{2/3}} \frac{dy_{1}}{y_{1}}
~U_{q}(t_{1},Ty_{1}^{3/2})
\nonumber
\ee
\be
\times\int_{(t_{0}/T)^{2/3}}^{(t_{2}/T)^{2/3}} ~\frac{dy_{2}}{y_{2}}
~U_{q}(t_{2},Ty_{2}^{3/2})
\nonumber
\ee
\be
\times H_{0}(Q\sqrt{y_{1}})
H_{0}(Q\sqrt{y_{2}})
\nonumber
\ee
\be
\times\tilde{\ell}^{d}(y_{1},y_{2})
F_{\sigma}(Q\tilde{\ell}(y_{1},y_{2}),Ty_{1}^{3/2},Ty_{2}^{3/2})
,
\label{eq:52}
\ee
and, with $\bar{y}_{1}=z_{1}^{2/3}$ in the integral,
\be
U_{q}(t_{1},Ty_{1}^{3/2})=e^{\left(-\int_{y_{1}}^{(t_{1}/T)^{2/3}}
\frac{d\bar{y}_{1}}{\bar{y}_{1}}H(Q\sqrt{\bar{y}_{1}})\right)}
~~~.
\ee
For our simple polynomial model given by Eq.(\ref{eq:31}) we can
carry out the $\bar{y}_{1}$ integration in $U_{q}$ explicitly.

\section{Auxiliary Field Scaling Function}

We must now work out the scaling properties of the field
$\sigma (m({\bf r},t))$ in the case where $m$ is a gaussian
variable.  It is well known that in the scaling regime
\be
C_{\sigma}({\bf r}_{1},t_{1},{\bf r}_{2}t_{2})
=<\sigma (m({\bf r}_{1},t_{1}))\sigma (m({\bf r}_{2},t_{2}))>
\nonumber
\ee
\be
=\psi_{0}^{2}\frac{2}{\pi}sin^{-1} ~f_{0}({\bf r}_{1},t_{1},{\bf r}_{2},t_{2})
~~~,
\label{eq:97}
\ee
where 
\be
f_{0}({\bf r}_{1},t_{1},{\bf r}_{2},t_{2})
=\frac{C_{0}({\bf r}_{1},t_{1},{\bf r}_{2},t_{2})}
{\sqrt{S_{0}(t_{1})S_{0}(t_{2})}}
\label{eq:83}
~~~,
\ee
\be
C_{0}({\bf r}_{1},t_{1},{\bf r}_{2},t_{2})
=<m({\bf r}_{1},t_{1})m({\bf r}_{2},t_{2})>
~~~,
\label{eq:A}
\ee
and
\be
S_{0}(t_{1})=C_{0}({\bf r}_{1},t_{1},{\bf r}_{1},t_{1})
~~~.
\label{eq:B}
\ee
Thus we need to focus on the determination of the auxiliary
field correlation function $C_{0}$.

Let us assume that the general equation of motion satisfied at
gaussian level by the auxiliary field $m$ has the local form in
Fourier space:
\be
\frac{\partial m_{q}(t)}{\partial t}=-\omega _{q}(t)  m_{q}(t)
\label{eq:77}
\ee
for $t > t_{0}$.  Since the growth law is of the form
$L=L_{0}t^{1/3}$, we see that for a scaling result we must
choose
$\omega _{q}(t)\approx {\cal O}\left(L^{-3}\right)$.
Since we can estimate  $q\approx {\cal O}\left( L^{-1}\right)$ in the scaling 
regime, we write quite generally that
\be
\omega _{q}(t)=\alpha_{0}\frac{1}{L^{3}}+\alpha_{1}\frac{q}{L^{2}}
+ \alpha_{2}\frac{q^{2}}{L}+\alpha_{3}q^{3}
\nonumber
\ee
\be
=\frac{1}{L^{3}}\left(\alpha_{0}+\alpha_{1}Q+\alpha_{2}Q^{2}
+\alpha_{2}Q^{2}+\alpha_{3}Q^{3}\right)
\label{eq:78}
\ee
where $Q=qL$.
We can use general arguments, similar to those due to Ohta and 
Nozaki\cite{ON},
to justify including the odd terms in $Q$.
Eq.(\ref{eq:78}) can be rewritten in the convenient form
\be
\omega _{q}(t) =\frac{\partial \Omega (Q(t))}{\partial t}
-\omega _{0}\frac{\partial }{\partial t}~ln~ t
~~~,
\label{eq:79}
\ee
where
\be
\Omega (Q)=a Q+bQ^{2}+cQ^{3}
~~~,
\label{eq:66}
\ee
$\omega_{0}=-\alpha_{0}/L_{0}^{3}$,
$a=\alpha_{1}/L_{0}^{2}$,
$b=\alpha_{2}/(2L_{0}^{2})$,
and $c=\alpha_{3}/(3L_{0}^{2})$.

The equation of motion for the auxiliary field,
Eq.(\ref{eq:77}),  has the general solution
\be
m_{q}(t)=e^{-\int_{t_{0}}^{t}d\tau~\omega _{q}(\tau )} m_{q}(t_{0})
~~~.
\ee
Inserting Eq.(\ref{eq:79}) into the integral in the exponential gives
\be
m_{q}(t)= e^{-(\Omega (Q(t))-\Omega (Q(t_{0}))}
e^{\omega _{0}~ln (t/t_{0})} m_{q}(t_{0})
\nonumber
\ee
\be
=\left(\frac{t}{t_{0}}\right)^{\omega_{0}}e^{-(\Omega (Q(t))-\Omega (Q(t_{0}))}
m_{q}(t_{0}) ~~~.
\ee
If we then form the auxiliary field correlation function,
as defined by the Fourier transform of Eq.(\ref{eq:A}), we obtain
\be
C_{0}(q,t_{1},t_{2})=\left(\frac{t_{1}t_{2}}{t_{0}^{2}}\right)^{\omega_{0}}
e^{-(\Omega (Q(t_{1}))+\Omega (Q(t_{2}))-2\Omega (Q(t_{0}))}
\nonumber
\ee
\be
\times C_{0}(q,t_{0},t_{0}) ~~~.
\label{eq:69}
\ee
If we go to the equal-time limit this becomes
\be
C_{0}(q,t,t)=\left(\frac{t}{t_{0}}\right)^{2\omega_{0}}
e^{-(2\Omega (Q(t))-2\Omega (Q(t_{0}))}
C_{0}(q,t_{0},t_{0}) ~~~.
\label{eq:74}
\ee
The important autocorrelation function, defined by Eq.(\ref{eq:B}),
is given then by
\be
S_{0}(t)=\int\frac{d^{d}q}{(2\pi)^{d}}C_{0}(q,t,t)
\nonumber
\ee
\be
=\frac{1}{L^{d}(t)}\int\frac{d^{d}Q}{(2\pi)^{d}}
 \left(\frac{t}{t_{0}}\right)^{2\omega_{0}}
e^{-(2\Omega (Q)-2\Omega (QL(t_{0})/L(t)))}
\nonumber
\ee
\be
\times C_{0}(Q/L(t),t_{0},t_{0})
~~~.
\label{eq:75}
\ee
For large $L(t)$ we can replace
\be
C_{0}(Q/L(t),t_{0},t_{0})\rightarrow g_{0}
~~~,
\ee
where $g_{0}$ is characteristic of the initial correlation function, and
\be
2\Omega (Q)-2\Omega (QL(t_{0})/L(t))\rightarrow 2\Omega (Q) ~~~,
\ee
to obtain in Eq.(\ref{eq:75})
\be
S_{0}(t)=\frac{g_{0}}{L^{d}(t)}\left(\frac{t}{t_{0}}\right)^{2\omega_{0}}
\int\frac{d^{d}Q}{(2\pi)^{d}}e^{-(2\Omega (Q)}
\nonumber
\ee
\be
=\frac{g_{0}}{L^{d}(t)}\left(\frac{t}{t_{0}}\right)^{2\omega_{0}}J_{d}
~~~.
\label{eq:78a}
\ee
Here we have introduced the constant
\be
J_{d}=\int\frac{d^{d}Q}{(2\pi)^{d}}e^{-2\Omega (Q)}
~~~.
\ee
By assumption the auxiliary field scales with the growth law,
$m\approx L$ and $S_{0}(t)\approx L^{2}(t)$.  Using this result
back in Eq.(\ref{eq:78a}) gives
\be
S_{0}(t)\approx L^{2\omega_{0}3-d}\approx L^{2}
~~~.
\ee
This fixes the constant $\omega_{0}$ to have the value:
\be
\omega_{0}=\frac{2+d}{6} ~~~.
\ee
Using this result in the scaling regime,
the auxiliary field correlation function, given by Eq.(\ref{eq:69})  
can be written 
in the form
\be
C_{0}(q,t_{1},t_{2})=\left(\frac{t_{1}t_{2}}{t_{0}^{2}}\right)^{\omega_{0}}
e^{-(\Omega (Q(t_{1}))+\Omega (Q(t_{2})))} g_{0}
~~~.
\ee
Eventually we need the inverse Fourier transform of the normalized 
correlation function
\be
f_{0}(q,t_{1},t_{2})=\frac{C_{0}(q,t_{1},t_{2})}{\sqrt{S_{0}(t_{1})S_{0}(t_{2})}}
~~~.
\ee
Using previous results this
can easily be put into the form
\be
f_{0}(q,t_{1},t_{2})=\left(L(t_{1})L(t_{2})\right)^{d/2}
\frac{e^{-(\Omega (Q(t_{1}))+\Omega (Q(t_{2})))}}{J_{d}}
~~~.
\label{eq:84}
\ee
With $\Omega (Q)$
given by Eq.(\ref{eq:66}), we have in the argument of the exponential:
\be
\Omega (Q(t_{1}))+\Omega (Q(t_{2}))=aq(L(t_{1})+L(t_{2}))
\nonumber
\ee
\be
+bq^{2}(L^{2}(t_{1})+L^{2}(t_{2}))
+cq^{3}(L^{3}(t_{1})+L^{3}(t_{2}))
~~~.
\ee
If we introduce $L_{T}(t_{1},t_{2})$, as in Eq.(\ref{eq:39}),
and $Q=qL_{T}$,  we obtain
\be
\Omega (Q(t_{1}))+\Omega (Q(t_{2}))=2aQ\ell_{1}(t_{1},t_{2})
\nonumber
\ee
\be
+2bQ^{2}\ell_{2}(t_{1},t_{2})
+2cQ^{3}\ell_{3}(t_{1},t_{2})
\ee
where
\be
\ell_{n}(t_{1},t_{2})
=\frac{1}{2}\left(\frac{L^{n}(t_{1})+L^{n}(t_{2})}{L_{T}^{n}(t_{1},t_{2})}\right)
\nonumber
\ee
\be
=\frac{1}{2}\frac{t_{1}^{n/3}+t_{2}^{n/3}}
 {\left(\frac{1}{2}\left(t_{1}^{2/3}+t_{2}^{2/3}\right)\right)^{n/2}}
~~~.
\ee
For equal times 
$\ell_{n}(t,t)= 1$,
while for $t_{1}\gg t_{2}$
\be
\ell_{n}(t_{1},t_{2})=2^{n/2 -1}
\ee
and
$\ell_{2}(t_{1},t_{2})=1$.
We can then choose $L_{0}$ such that $c=1/4$, $2b=\mu$ and $2a=-A$.
We then have the final result
\be
f_{0}(q,t_{1},t_{2})=\left(L(t_{1})L(t_{2})\right)^{d/2}
\frac{e^{-N(Q,t_{1},t_{2})}}{J_{d}}
\label{eq:93}
\ee
where
\be
N(Q,t_{1},t_{2})=\frac{1}{2}Q^{3}\ell_{3}(t_{1},t_{2})+\mu Q^{2}-AQ\ell_{1}(t_{1},t_{2})
~~~.
\ee
The inverse Fourier transform of $f_{0}(q,t_{1},t_{2})$ is the quantity which
is related to the
$\sigma$- correlation function by Eq.(\ref{eq:97}):
\be
f_{0}(r,t_{1},t_{2})=\int \frac{d^{d}q}{(2\pi)^{d}}e^{-i\vec{q}\cdot\vec{r}}
f_{0}(q,t_{1},t_{2})
\nonumber
\ee
\be
=\int \frac{d^{d}q}{(2\pi)^{d}}e^{-i\vec{q}\cdot\vec{r}}
\left(L(t_{1})L(t_{2})\right)^{d/2}
\frac{e^{-N(Q,t_{1},t_{2})}}{J_{d}}
\nonumber
\ee
\be
=\left(\frac{L(t_{1})L(t_{2})}{L_{T}^{2}(t,t_{2})}\right)^{d/2}
\int \frac{d^{d}Q}{(2\pi)^{d}}e^{-i\vec{Q}\cdot\vec{x}}
\frac{e^{-N(Q,t,t_{2})}}{J_{d}}
\ee
where $x=r/L_{T}$ and now
\be
J_{d}=\int \frac{d^{d}Q}{(2\pi)^{d}}e^{-N(Q,t,t)}
\nonumber
\ee
\be
=\int \frac{d^{d}Q}{(2\pi)^{d}}e^{-\frac{1}{2}Q^{3}-\mu Q^{2}+AQ}
~~~.
\ee
The Fourier transform of the $\sigma$-correlation function
is given by:
\be
C_{\sigma}(q,t_{1},t_{2})=\int d^{d}~re^{+i\vec{q}\cdot\vec{r}}~
\frac{2}{\pi}sin^{-1}\left(f_{0}(r,t_{1},t_{2})\right)
\nonumber
\ee
\be
=L_{T}^{d}\int d^{d}x~ e^{+i\vec{Q}\cdot\vec{x}}~
\frac{2}{\pi}sin^{-1}\left(f_{0}(x,t_{1},t_{2})\right)
\ee
and the scaling function appearing in our physical scaling
function, Eq.(\ref{eq:97}), is given by
\be
F_{\sigma}(Q,t_{1},t_{2})=\int d^{d}x~ e^{+i\vec{Q}\cdot\vec{x}}~
\frac{2}{\pi}sin^{-1}\left(f_{0}(x,t_{1},t_{2})\right)
\label{eq:115}
\ee
where
\be
f_{0}(x,t_{1},t_{2})=
\left(\frac{L(t_{1})L(t_{2})}{L_{T}^{2}(t_{1},t_{2})}\right)^{d/2}
\nonumber
\ee
\be
\times\int \frac{d^{d}Q}{(2\pi)^{d}}e^{-i\vec{Q}\cdot\vec{x}}
~\frac{e^{-N(Q,t_{1},t_{2})}}{J_{d}}
~~~.
\label{eq:92}
\ee

\section{Large $Q$ and Short-distances}

One of the main successes of the OJK theory is inclusion of the
short distance nonanalytic behavior associated with Porod's
law and the Tomita sum rule.  We look here at how all of this
fits into the present development.

Let us consider Eq.(\ref{eq:52}) for equal times $t_{1}=t_{2}=t\gg t_{0}$
where
\be
F(Q)=F(Q,t,t)=
\int_{0}^{1} \frac{dy_{1}}{y_{1}}
e^{-\int_{y_{1}}^{1}\frac{d\bar{y}_{1}}{\bar{y}_{1}}H(Q\sqrt{\bar{y}_{1}})}
\nonumber
\ee
\be
\times\int_{0}^{1} ~\frac{dy_{2}}{y_{2}}
e^{-\int_{y_{2}}^{1}\frac{d\bar{y}_{2}}{\bar{y}_{2}}H(Q\sqrt{\bar{y}_{2}})}
\times H_{0}(Q\sqrt{y_{1}})
H_{0}(Q\sqrt{y_{2}})
\nonumber
\ee
\be
\tilde{\ell}^{d}(y_{1},y_{2})
F_{\sigma}(Q\tilde{\ell}(y_{1},y_{2}),ty_{1}^{3/2},ty_{2}^{3/2})
~~~.
\label{eq:93a}
\ee
At first sight, since $H(Q\sqrt{\bar{y}_{1}})$ goes at least as fast as
$Q^{2}$ for large $Q$, this integral looks like it gives exponential
behavior in $Q$ for large $Q$.  Closer inspection shows the 
asymptotic dependence on $Q$ is much slower in those regions of 
the $y_{1}$ and $y_{2}$
integrals near $1$.  By expanding $\tilde{\ell}^{d}(y_{1},y_{2})
F_{\sigma}(Q\tilde{\ell}(y_{1},y_{2}),ty_{1}^{3/2},ty_{2}^{3/2})$ about
$y_{1}=1$ and $y_{2}=1$, it is not difficult to show,
assuming that $H_{0}$ and $H$ increase algebraically with
$Q$ for large $Q$, that
\be
F(Q)=F_{\sigma}(Q)\frac{H_{0}(Q)}{H(Q)} (1+{\cal O}\left(H(Q)\right)^{-1})
~~~.
\ee
We will now show that  $F_{\sigma}(Q)$ falls off algebraically
with large wavenumber.  This means that the short-distance behavior
is controlled, as in the OJK theory, by the auxiliary field $m$.  We see that
the cross-over functions $H$ and $H_{0}$ do not influence this
short-distance behavior if we choose $H=H_{0}$.  Notice also
that it is advantageous to choose $H(Q)$ to have a high-power of
$Q$ component so one can ignore the corrections  to $F(Q)=F_{\sigma}(Q)$.
For this reason we keep the $\gamma_{10}$ term in Eq.(\ref{eq:31}) and the
crossover function $H(Q)$ does not influence the large $Q$ behavior of
$F(Q)$ until terms of ${\cal O}(Q^{-14})$.  We do not expect 
significant variations in our numerical results to depend on whether we keep,
for example,
$\gamma_{10}$ or $\gamma_{8}$.

Let us look in more detail at the large $Q$ behavior of $F_{\sigma}(Q)$.
This requires several steps.  First we have, using Eq.(\ref{eq:97}), that in coordinate
space
\be
F_{\sigma}(x)=\frac{2}{\pi}sin^{-1} f_{0}(x)
\label{eq:95}
\ee
where $f_{0}(x)$ is given by Eq.(\ref{eq:92}) with
$t_{1}=t_{2}=t$:
\be
f_{0}(x)=\int~\frac{d^{d}Q}{(2\pi )^{d}}~e^{-i{\bf Q}\cdot{\bf x}}
\frac{e^{-N(Q)}}{J_{d}}
~~~.
\label{eq:96a}
\ee
The short-scaled distance
expansion for $f_{0}(x)$ 
can be written in the form
\be
f_{0}(x)=1-a_{0}x^{2}\left(1+b_{0}x^{2}+c_{0}x^{4}+d_{0}x^{6}+\cdots\right)
~~~.
\label{eq:96}
\ee
If we define the integrals
\be
I_{n}=\int_{0}^{\infty}dQ~Q^{n} e^{-N(Q)}
\ee
then
\be
a_{0}=\frac{1}{6}\frac{I_{4}}{I_{2}}
\ee
\be
b_{0}=-\frac{1}{20}\frac{I_{6}}{I_{4}}
\ee
\be
c_{0}=\frac{6}{7!}\frac{I_{8}}{I_{4}}
\ee
\be
d_{0}=-\frac{6}{9!}\frac{I_{10}}{I_{4}}
~~~.
\ee
Inserting Eq.(\ref{eq:96}) into Eq.(\ref{eq:95}) and expanding, it is only a matter
of stamina to show:
\be
F_{\sigma}(x)=1-\alpha x\left (1+\beta x^{2}
+\gamma x^{4}+\nu  x^{6}+\cdots\right)
\label{eq:102}
\ee
where
\be
\alpha =\frac{2}{\pi}\sqrt{2a_{0}}
\ee
\be
\beta =\frac{a_{0}}{12}+\frac{b_{0}}{2}
\ee
\be
\gamma =\frac{c_{0}}{2}-\frac{b_{0}^{2}}{8}+\frac{a_{0}b_{0}}{8}
+\frac{3}{160}a_{0}^{2}
\ee
\be
\nu=\frac{d_{0}}{2}+\frac{a_{0}^{3}}{7!}-\frac{4}{105}\beta a_{0}^{2}
\nonumber
\ee
\be
+\frac{25}{56}\beta^{2} a_{0}+\frac{9}{28}\gamma a_{0}-\gamma\beta
~~~.
\ee
Thus all of the expansion coefficients for $F_{\sigma}(x)$ are known  in terms of the parameters
$A$ and $\mu$ in $N(Q)$.  

The large-Q behavior of $F_{\sigma}(Q)$  follows in three dimensions from the
Fourier representation
\be
F_{\sigma}(Q)=\frac{4\pi}{Q}\int_{0}^{\infty}dx~x~sin(Qx)F_{\sigma}(x)
\ee
and, after repeated integrations by parts, one obtains for large
$Q$
\be
F_{\sigma}(Q)=\sum_{n=2}^{\infty}\frac{F_{2n}}{Q^{2n}}
\ee
where the Porod coefficients are defined by:
\be
F_{2n}=4\pi\left(-1\right)^{n+1}
\left(\frac{d^{2n-2}}{dx^{2n-2}}(xF_{\sigma}(x))\right)_{x=0}
~~~.
\label{eq:109}
\ee
Using Eq.(\ref{eq:102}) in Eq.(\ref{eq:109}) we easily find
\be
F_{4}=8\pi\alpha
\ee
\be
F_{6}=-96\pi\alpha\beta
\ee
\be
F_{8}=4\pi 6!\alpha\gamma
\ee
\be
F_{10}=-4\pi 8!\alpha\nu
~~~.
\ee

Following Tomita we can show that the lack of even terms in the
expansion of $F(x)$ and $F_{\sigma}(x)$ leads to the set of sum rules
\be
S_{n}=\int\frac{d^{3}Q}{(2\pi)^{3}}Q^{2n}
\left[F(Q)-\sum_{m=2}^{n+1}\frac{F_{2m}}{Q^{2m}}\right] =0 ~~~.
\label{eq:140}
\ee
If we set $F(Q)=F_{\sigma}(Q)$ in Eq.(\ref{eq:140}), because of
Eq.(\ref{eq:102}),  the sum rules
are obeyed identically.  However, if we insert $F(Q)$ given by
Eq.(\ref{eq:93a})  into Eq.(\ref{eq:140}), there is no reason in general to
expect the sum rules to be satisfied.  Indeed this gives a set of
conditions on $F(Q)$ which can be used, along with the normalization
\be
S_{0}=\int \frac{d^{3}Q}{(2\pi )^{3}}~F(Q) -1 =0 ~~~,
\label{eq:136}
\ee
to determine the parameters $A$, $\mu$ and those determining
$H(Q)$.

\section{Determination of the scaling function}

The equal-time scaling function $F(Q)$ given by Eq.(\ref{eq:52}) is a 
function of the $\gamma_{n}$, $\gamma_{n}^{0}$, $A$ and $\mu$.
Our basic assumption is that our model is characterized by the 
normalization Eq.(\ref{eq:136}) and the set of sum rules given by
Eq.(\ref{eq:140}).  Thus we have an infinite set of parameters
and an infinite set of conditions.  Here we work out the
truncated theory where  we use the four
conditions $S_{i}=0$, for $i=0,1,2,3$ to determine the parameters
$P=\{\gamma_{2},\gamma_{10}, A, \mu\}$
with $\gamma_{n}=\gamma_{n}^{0}$.
Thus there are no free parameters.

One then has a rather complicated numerical minimization problem.
First assume values of the four parameters  $P^{(1)}$ and
compute $F^{(1)}(Q)$ and simultaneously the four sum rules
$S_{n}^{(1)}$.  Then choose another set $P^{(2)}$ and 
determines the set $S_{n}^{(2)}$.  One then needs a measure, like
\be
{\cal J}^{(i)}=\sum_{n=0}^{3}\left( S_{n}^{(i)}\right)^{2}
~~~,
\ee
to minimize.  Thus, if ${\cal J}^{(2)}<{\cal J}^{(1)}$, the
set of parameters $P^{(2)}$ is preferred over the set $P^{(1)}$.
One then iterates this process until it converges to a selected
{\it fixed-point} set of  values of the parameters $P^{*}$ and
scaling function $F^{*}(Q)$.

In the numerical determination of the $S_{n}$ it is important to take into account that 
these integrals are slowly converging for large wavenumbers. Suppose we have  
integrated a sum rule out to a cutoff wavenumber $Q_{M}$  and obtain a contribution
$S_{n}^{Q_{M}}$.  If $Q_{M}$ is large enough, the system is dominated 
by the large-$Q$ power-law
behavior as in Porod's law and one can evaluate the remaining contribution
to the integral in terms of an appropriate Porod coefficient:
\be
S_{n}=S_{n}^{Q_{M}}+\frac{F_{2n+2}}{2\pi Q_{M}}
~~~.
\label{eq:114}
\ee

The determination of the $S_{n}^{Q_{M}}$ is a challenging numerical problem
since one must perform
multiple nested integrations $y_{1},y_{2}$ and internal Fourier transforms
and still maintain sufficient numerical accuracy that the Porod
coefficients $F_{2n}$ can be extracted  and the sum
rules constructed.  Thus the choice $Q_{M}$ must be monitored
carefully when determining $S_{n}^{Q_{M}}$.

As a result of extensive iterations we arrive at the fixed point
values\cite{loosen} for the parameters: $\gamma_{2}=1.4994\ldots$, 
$\gamma_{10}=0.1232\ldots$, $A=6.395\ldots$, and $\mu=0.2162\ldots$.
In Fig. 1 we plot the determined scaling function $F(Q)$ versus the
accurate numerical results of Oono and Shinozaki\cite{SO}.  Notice
that the structure factor is normalized by the position and height of
the first maximum.  The agreement for small $Q$ is very good and the width
of the peak is in good agreement with the numerical results.  In Fig.2
we plot $Q^{4}F(Q)$ for the theory and the same set of numerical results.
Again overall agreement is good.  The value of the Porod coefficient $F_{4}$
for the theory and simulation are in excellent agreement.  The feature
of a second maximum is present in the theory but its position and width
are not in very good agreement with the numerical results.  In Figs.3-6
we plot the running contributions to the sum rules to give a feeling for the degree of
convergence of the numerical procedure described above.  To obtain the
complete contribution to the sum rules we must add the last term in
Eq.(\ref{eq:114}).  The analytically determined Porod coefficients needed
in completing the analysis are given by $F_{4}=21.87$, $F_{6}=19.72$,
$F_{8}=-52.85$ and $F_{10}=-25.41$.

The results here should be compared with previous direct calculations of the
$F(Q)$.  In Ref.(\onlinecite{TUGCOP}) the calculated $F(Q)$ obeyed Porod's law, did not obey
the lowest Tomita sum rule, went as $Q^{2}$ for small $Q$ and gave a
width significantly too broad compared to numerical results.  In 
Ref.(\onlinecite{GFMCOP}) the large and small $Q$ limits were in agreement with
expectations but the Tomita sum rule was not satisfied and the width
of the scaling function was significantly {\it too ~narrow}.  In
Ref.(\onlinecite{ON}) the authors obtained a scaling function which satisfied
Porod's law, the Tomita sum rules and was fit to the numerically determined
width.  Unfortunately, this work not only did not give the $Q^{4}$ small
$Q$ behavior for $F(Q)$, it did not satisfy the conservation law,
$F(0)>0$.

\section{Autocorrelation Function Exponents}

We turn next to a discussion of the two-time order parameter correlation function.
In this case we focus on the
behavior of the autocorrelation function, which, when
both $t_{1}$ and $t_{2}$ are in
the scaling regime, is given by
\be
\Psi (t_{1},t_{2})=<\psi ({\bf r},t_{1})\psi ({\bf r},t_{2})>
\nonumber
\ee
\be
=\int ~\frac{d^{d}q}{(2\pi )^{d}}~ C(q,t_{1},t_{2})
=\psi_{0}^{2}\int ~\frac{d^{d}Q}{(2\pi )^{d}}~ F(Q,t_{1},t_{2})
\nonumber
\ee
\be
=\psi_{0}^{2}\int ~\frac{d^{d}Q}{(2\pi )^{d}}~
\int_{(t_{0}/T)^{2/3}}^{(t_{1}/T)^{2/3}} \frac{dy_{1}}{y_{1}}
~U_{q}(t_{1},Ty_{1}^{3/2})
\nonumber
\ee
\be
\times\int_{(t_{0}/T)^{2/3}}^{(t_{2}/T)^{2/3}} ~\frac{dy_{2}}{y_{2}}
~U_{q}(t_{2},Ty_{2}^{3/2})
H_{0}(Q\sqrt{y_{1}})
H_{0}(Q\sqrt{y_{2}})
\nonumber
\ee
\be
\times\ell^{d}(y_{1},y_{2})
F_{\sigma}(Q\ell(y_{1},y_{2}),Ty_{1}^{3/2},Ty_{2}^{3/2})
~~~.
\label{eq:138}
\ee
In the regime $t_{1}\gg t_{2}$, $\Psi (t_{1},t_{2})$ 
is expected to take
the form\cite{221}
\be
\Psi (t_{1},t_{2})\approx \left(\frac{L(t_{2})}{L(t_{1})}\right)^{\lambda}
\ee
and the exponent $\lambda$ is distinct from
the growth law.  It is not difficult to extract $\lambda$ from
Eq.(\ref{eq:138}).

For $t_{1}\gg t_{2}$ the integral over $y_{2}$ is restricted to
small values and the autocorrelation function can be put into the
form
\be
\Psi (t_{1},t_{2})=\psi_{0}^{2}\int ~\frac{d^{d}Q}{(2\pi )^{d}}~
\int_{0}^{2} \frac{dy_{1}}{y_{1}}
e^{\left(-\int_{y_{1}}^{2}
\frac{d\bar{y}_{1}}{\bar{y}_{1}}H(Q\sqrt{\bar{y}_{1}})\right)}
\nonumber
\ee
\be
\times H_{0}(Q\sqrt{y_{1}})
\int_{0}^{2(t_{2}/t_{1})^{2/3}} ~\frac{dy_{2}}{y_{2}}
\gamma_{2}Q^{2}y_{2}
\nonumber
\ee
\be
\times \tilde{\ell}^{d}(y_{1},0)
F_{\sigma}(Q\tilde{\ell}(y_{1},0),Ty_{1}^{3/2},Ty_{2}^{3/2})
~~~~.
\label{eq:140a}
\ee
They key observation is that for $t_{1}\gg t_{2}$ the auxiliary
field correlation function is small and we can linearized  
Eq.(\ref{eq:115}) relating
the order parameter and the auxiliary
field correlation function to obtain
\be
F_{\sigma}(Q,t_{1},t_{2})\approx \frac{2}{\pi}f_{0}(Q,t_{1},t_{2})
~~~,
\ee
where
\be
f_{0}(Q,t_{1},t_{2})
=\left(\frac{L(t_{1})L(t_{2})}{L_{T}^{2}}\right)^{d/2}
\frac{e^{-N(Q,t_{1},t_{2})}}{I_{d}}
~~~.
\ee
Then to leading order for $t_{1}\gg t_{2}$
\be
f_{0}(Q,t_{1},t_{2})
\nonumber
\ee
\be
=\left(\frac{2 L(t_{2})}{L(t_{1})}\right)^{d/2}
\frac{e^{-\frac{1}{\sqrt{2}}\left(Q^{3}+\sqrt{2}\mu Q-AQ\right)}}{I_{d}}
\label{eq:A1}
~~~.
\ee
Inserting this result back
into Eq.(\ref{eq:140a}), we find that the overall time dependence is governed
by the $y_{2}$ integral given by
\be
\int _{0}^{2(L(t_{2})/L(t_{1}))^{2}}dy_{2}~y_{2}^{d/4}
\approx (L(t_{2})/L(t_{1}))^{d/2+2}
\ee
and we can  identify\cite{lower}  $\lambda =d/2+2$.
This result corresponds to the lower bound established
by Yeung, Rao and Desai\cite{YRD}.

If $t_{2}=t_{0}$ then the analysis must be altered.  The key point
is that one must keep the first term in Eq.(\ref{eq:29}) and the dominant
term for long times is the correlation between the second term in
Eq.(\ref{eq:29}) for the field at $t_{1}$ and the initial correlation.  Then
one has
\be
C(q,t_{1},t_{0})=\int_{t_{0}}^{t_{1}}~d\bar{t}~U_{q}(t_{1},\bar{t})
N_{q}(\bar{t})<\sigma_{q}(\bar{t})\sigma_{-q}(\bar{t_{0}})>
\ee
The analysis of this quantity for large $t_{1}$ follows the earlier
analysis with respect to rescaling
of the $t_{1}$ dependence.  The main difference in the calculation
is that there is no $y_{2}$ integration as in the
$t_{2}\gg t_{0}$ case
and one has the direct factor as in Eq.(\ref{eq:A1}) which leads
to the behavior
\be
\Psi (t_{1},t_{2})\approx (L_{1})^{-d/2}
~~~.
\ee
In this case we again 
obtain that the exponent is given by
the lower bound value $\lambda_{0} =d/2$ found by Yeung, et al\cite{YRD}.

\section{Discussion}

In this paper we have developed a theory of the phase-ordering
kinetics of the CH model which plays a role similar to
the OJK theory for the NCOP case.  The theory includes all of the desired
features discussed in section 2 including the elusive Tomita sum rule.  It is
also in good numerical agreement  with the best available numerical results. 
The theory, because of the nonlocal
mapping onto the auxiliary field, is much more difficult to treat 
analytically when compared to the OJK theory.  Nevertheless it seems to work
well.  While we are able to extract
the apparently mean field values for the nonequilibrium exponents
$\lambda$ and $\lambda_{0}$, it looks very difficult to include the 
nonlinear terms in the $m$ field necessary to give
the higher-order corrections for the exponents.

There are several directions in which this work can be extended.
One can, in principle, include additional $\gamma_{n}$ parameters 
in the model and satisfy
higher order sum rules.  This will be numerically difficult and
probably will not improve the scaling function significantly.
While the detailed analysis here has been for three dimensions, with
only a modest amount of additional analytical work, the selection process
can be applied to two dimensional systems.
The most interesting new direction is to apply this same theory to off
critical quenches.  The expectation is that the scaled structure factor
broadens significantly as one moves away from critical quenches.  It will be
interesting to see how well the theory describes the system as one moves toward
the coexistence curve and the regime where the Lifshitz-Slyosov-Wagner theory\cite{Ohta}
is applicable.

\section{Acknowledgement}

This work supported in part by the MRSEC Program of the National
Science Foundation under Contract No. DMR-9808595.

\onecolumn

\begin{figure}
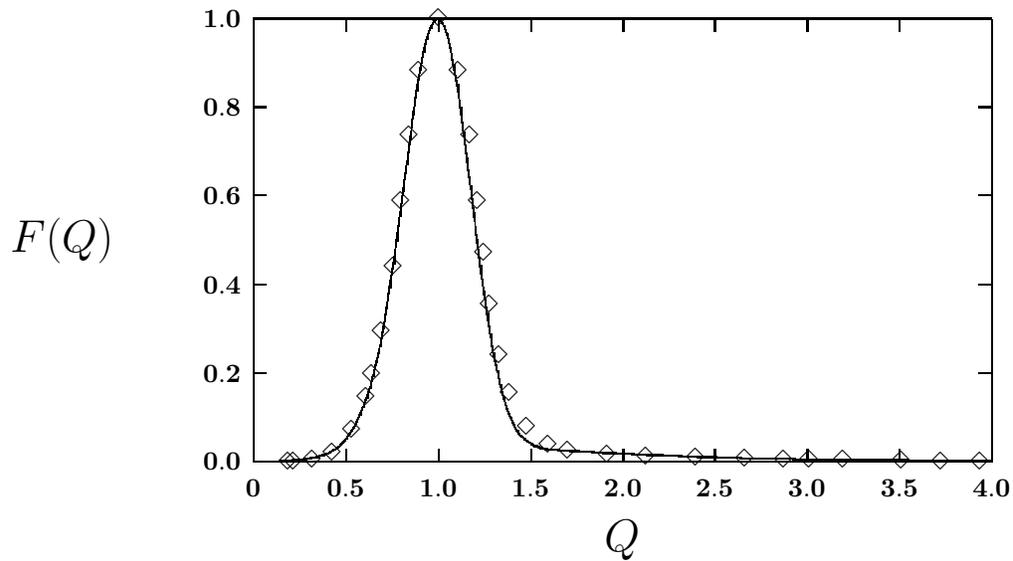


\setlength{\unitlength}{0.240900pt}
\ifx\plotpoint\undefined\newsavebox{\plotpoint}\fi
\sbox{\plotpoint}{\rule[-0.200pt]{0.400pt}{0.400pt}}%


\caption{Plot of scaled structure factor versus scaled wavenumber.
Solid line represent the theory and the squares the numerical
results from Ref.([22])}

\end{figure}

\begin{figure}

\setlength{\unitlength}{0.240900pt}
\ifx\plotpoint\undefined\newsavebox{\plotpoint}\fi
\sbox{\plotpoint}{\rule[-0.200pt]{0.400pt}{0.400pt}}%
%

\caption{Porod plot of scaled structure factor versus scaled wavenumber.
Solid line represent the theory and the squares the numerical
results from Ref.([22]}

\end{figure}

\begin{figure}
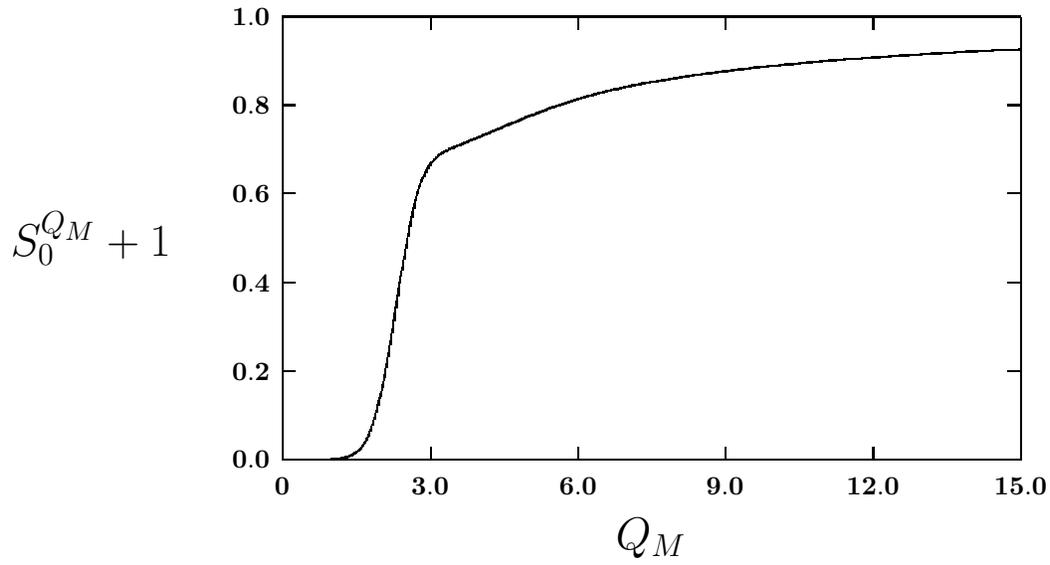


\setlength{\unitlength}{0.240900pt}
\ifx\plotpoint\undefined\newsavebox{\plotpoint}\fi
\sbox{\plotpoint}{\rule[-0.200pt]{0.400pt}{0.400pt}}%
%

\caption{Plot of normalization of structure factor versus cutoff wavenumber}

\end{figure}

\begin{figure}

\setlength{\unitlength}{0.240900pt}
\ifx\plotpoint\undefined\newsavebox{\plotpoint}\fi
\sbox{\plotpoint}{\rule[-0.200pt]{0.400pt}{0.400pt}}%
%

\caption{Plot of sum rule $S_{1}^{Q_{M}}$ versus cutoff wavenumber}

\end{figure}

\begin{figure}

\setlength{\unitlength}{0.240900pt}
\ifx\plotpoint\undefined\newsavebox{\plotpoint}\fi
\sbox{\plotpoint}{\rule[-0.200pt]{0.400pt}{0.400pt}}%
%

\caption{Plot of sum rule $S_{2}^{Q_{M}}$ versus cutoff wavenumber}

\end{figure}

\begin{figure}

\setlength{\unitlength}{0.240900pt}
\ifx\plotpoint\undefined\newsavebox{\plotpoint}\fi
\sbox{\plotpoint}{\rule[-0.200pt]{0.400pt}{0.400pt}}%
%

\caption{Plot of sum rule $S_{3}^{Q_{M}}$ versus cutoff wavenumber}

\end{figure}


\begin{thebibliography}{99}
\bibitem{few}  There are not very many possibilites for quantities which are
or ${\cal O}(1)$ if they are to be built out of $\sigma$ and $\psi$.  If
one allows gradients then one has quantities like $\left(\nabla m\right)^{2}$
as discussed in Ref.(\onlinecite{EXP}).
\bibitem{NL} This new mapping, unlike the case discussed in Ref.(\onlinecite{YOS}),
is nonlocal in both space and time.
\bibitem{loosen}  In order to speed up the convergence we have allowed the
ratio $r=H_{0}(Q)/H(Q)$ to float.  It takes on the fixed point value $1.0055\ldots$
in agreement with the assumption that it can be chosen to be 1.
\bibitem{lower}  The integration over $y_{1}$ is well behaved at its
lower limit if $d<4$.
\end{thebibliography}
\end{document}